\documentclass[11pt,preprint]{aastex}
\usepackage{bm}
\usepackage{epsfig}

\begin{document}

\title{\bf Gravitational convergence, shear deformation and rotation of magnetic forcelines}

\author{Vangelis Giantsos\footnote{Current Address: Faculty of Physics, Theoretical Physics, Ludwig Maximilians University, Theresienstrasse 37, 80333 Munich, Germany.}~ and Christos G. Tsagas}

\affil{Section of Astrophysics, Astronomy and Mechanics, Department of Physics,\\ Aristotle University of Thessaloniki, Thessaloniki 54124, Greece}

\date{\empty}

\begin{abstract}
We consider the ``kinematics'' of spacelike congruences and apply them to a family of self-gravitating magnetic forcelines. Our aim is to investigate the convergence and the possible focusing of these lines, as well as their rotation and shear deformation. In so doing, we introduce a covariant 1+2 splitting of the 3-dimensional space, parallel and orthogonal to the direction of the field lines. The convergence, or not, of the latter is monitored by a specific version of the Raychaudhuri equation, obtained after propagating the spatial divergence of the unit magnetic vector along its own direction. The resulting expression shows that, although the convergence of the magnetic forcelines is affected by the gravitational pull of all the other sources, it is unaffected by the field's own gravity, irrespective of how strong the latter is. This rather counterintuitive result is entirely due to the magnetic tension, namely to the negative pressure the field exerts parallel to its lines of force. In particular, the magnetic tension always cancels out the field's energy-density input to the Raychaudhuri equation, leaving the latter free of any direct magnetic-energy contribution. Similarly, the rotation and the shear deformation of the aforementioned forcelines are also unaffected by the magnetic input to the total gravitational energy. In a sense, the magnetic lines do not seem to ``feel'' their own gravitational field no matter how strong the latter may be.
\end{abstract}

\section{Introduction}\label{sI}
The Raychaudhuri equation is a fully geometrical expression that has been traditionally used to monitor the convergence (or not) of timelike worldlines in relativistic studies of gravitational collapse (see~\cite{W,Po}), or the mean expansion of cosmological spacetimes (see~\cite{R1,D1,Eh,El3,D2} for representative discussions). Nevertheless, Raychaudhuri's formula is not a priori restricted to timelike curves and to 4-dimensional spacetimes. The same is also true for the supplementary equations monitoring the other two ``optical scalars'' (as they have been historically known -- e.g.~see~\cite{K}), namely the shear and the vorticity. Instead, generalised analogues of the Raychaudhuri equations (and of the rest of the supplementary formulae) can be used to study curves of any nature in diverse environments (e.g.~see~\cite{KS,AV}). For example, in astrophysics and cosmology~\citep{DNK1,DNK2,TK}, within quantum and modified-gravity scenarios~\citep{GND,HL,Da,Mo,PTB}, as well as in spaces of arbitrary dimensions~\citep{KUNZ,PNG}. The forcelines of a magnetic ($B$) field are spacelike curves, as seem by an observer at rest relative to them. Therefore, confining to the 3-dimensional rest-space of such an observer, one can construct the associated Raychaudhuri equation and the evolution formulae of the remaining optical scalars, to study the convergence (or not) of these lines, their shear deformation and their rotation.

Magnetic fields are rather unique sources and one of their special features is their tension properties. These reflect the elasticity of the magnetic forcelines, which is manifested as negative pressure exerted along the direction of the $B$-field~\citep{Pa,Me}. Our aim is to investigate the implications of this unique magnetic property for the ``kinematic'' behaviour of the field lines themselves. In so doing, we will adopt the so-called 1+1+2 covariant approach to general relativity~\citep{G,TM,MT,Z,CB,C}. The latter starts by introducing an 1+3 splitting of the spacetime, into time and 3-dimensional space, before proceeding to an additional 1+2 decomposition of the spacelike hypersurfaces along a given direction and 2-dimensional surfaces orthogonal to it. This preferred spatial direction also defines a unitary spacelike vector parallel to it. The ``kinematics'' of such a vector field, namely whether its (spacelike) tangent curves converge/diverge, rotate or change shape, are determined by a set of ``propagation'' equations analogous to those of their timelike counterparts~\citep{El1,El2}. The difference is that, here, the propagation is along a spatial direction, instead of a temporal one. For instance, the convergence/divergence of these spacelike curves (along their own direction), is monitored by the associated Raychaudhuri formula. In our study, it is the magnetic forcelines that single-out a preferred spatial direction and, in so doing, they also define a unit vector field parallel to them. Then, the associated Raychaudhuri equation and the rest of the propagation formulae determine whether (and under what conditions) these forcelines converge or diverge, whether they rotate relative to each other and whether their shape is deformed.

Perhaps the main difference between the kinematic equations of timelike worldlines and those of a spacelike congruence, is in their curvature terms. The former involve the Riemann and the Ricci tensors of the whole spacetime, while the latter their 3-dimensional (spatial) counterparts. In empty and static spaces all these tensors vanish identically, but in any other case they differ (sometimes considerably). For our purposes, the key difference appears to come from the pressure contribution. More specifically, although the isotropic pressure of the matter adds to the spacetime Ricci tensor, it does not contribute to its spatial analogue. The anisotropic (trace-free) pressure, on the other hand, does. This means that only the magnetic energy density and the anisotropic pressure contribute to the Raychaudhuri equation of the field lines. This brings into play the magnetic tension, which manifests itself as negative pressure in the direction of the $B$-field. What is important is that the tension contribution to the Raychaudhuri equation always cancels out the input of the magnetic energy density. As a result, the convergence or not of the field lines is not directly affected by their own gravitational energy, no matter how strong the latter may be. The same is also true for the rotation and the shear deformation of these lines. Overall, although the magnetic forcelines respond to the gravitational pull of all the other sources, they do not seem to ``feel'' their own gravity and this counterintuitive behaviour is exclusively due to their tension properties. This means that a magnetic-line configuration that finds itself at rest in an otherwise empty and static space will remain in equilibrium indefinitely, unless an external agent intervenes.

\section{Spacetime decomposition}\label{sSD}
Introducing a timelike 4-velocity field into the 4-dimensional spacetime achieves an 1+3 decomposition of the latter into a temporal direction and a 3-dimensional space orthogonal to it. In addition, selecting a spacelike direction and then decomposing the spatial sections parallel and orthogonal to it, leads to the so-called 1+1+2 spacetime splitting (see~\cite{G,TM,MT,Z}, as well as~\cite{CB} and~\cite{C}).

\subsection{1+3 splitting}\label{ss1+3S}
In a 4-dimensional spacetime, with metric $g_{ab}$, introduce a temporal direction along the timelike 4-velocity $u_a$ (normalised so that $u_au^a=-1$). Then, the symmetric tensor $h_{ab}=g_{ab}+u_au_b$ projects into the 3-dimensional (spatial) hypersurfaces orthogonal to $u_a$ (i.e.~$h_{ab}u^b=0$ with $h_{ab}h^b{}_c= h_{ac}$ and $h_a{}^a=3$). The 3-dimensional Levi-Civita tensor is $\varepsilon_{abc}=\eta_{abcd}u^d$ (where $\eta_{abcd}$ is its 4-dimensional counterpart) and satisfies the conditions $\varepsilon_{abc}=\varepsilon_{[abc]}$ and $\varepsilon_{abc}\varepsilon^{dmf}= 3!h_{[a}{}^dh_b{}^mh_{c]}{}^f$. All these allow for an 1+3 splitting of the spacetime into time and 3-dimensional space, parallel and orthogonal to $u_a$ respectively~\citep{TCM}. Then, the temporal and spatial derivatives of a general tensor field $T_{ab\cdots}{}^{cd\cdots}$ are given by
\begin{equation}
\dot{T}_{ab\cdots}{}^{cd\cdots}= u^s\nabla_sT_{ab\cdots}{}^{cd\cdots} \hspace{10mm} {\rm and} \hspace{10mm} {\rm D}_sT_{ab\cdots}{}^{cd\cdots}= h_s{}^qh_a{}^fh_b{}^k\cdots h_p{}^ch_r{}^d\cdots\nabla_qT_{fk\cdots}{}^{pr\cdots}\,,  \label{1+3dirs}
\end{equation}
respectively. Applying the above to the 3-dimensional projector ($h_{ab}$), leads to
\begin{equation}
\dot{h}_{ab}= 2u_{(a}\dot{u}_{b)} \hspace{10mm} {\rm and} \hspace{10mm} {\rm D}_ch_{ab}= 0\,,  \label{hdirs}
\end{equation}
the second of which shows why $h_{ab}$ can be used as the metric tensor of the spatial sections (in the absence of rotation -- see~\cite{TCM}).

Let us now consider a congruence of timelike worldlines tangent to the 4-velocity field $u_a$. Using definitions (\ref{1+3dirs}a) and (\ref{1+3dirs}b), we arrive at the decomposition~\citep{TCM}
\begin{equation}
\nabla_bu_a= {1\over3}\,\Theta h_{ab}+ \sigma_{ab}+ \omega_{ab}- \dot{u}_au_b\,.  \label{Nbua}
\end{equation}
On the right-hand side we have the irreducible kinematic variables of the congruence's motion. In particular, $\Theta=\nabla_au^a={\rm D}_au^a$ is the volume scalar, $\sigma_{ab}={\rm D}_{\langle b} u_{a\rangle}$ is the shear tensor, $\omega_{ab}={\rm D}_{[b} u_{a]}$ is the vorticity tensor and $\dot{u}_a=u^b\nabla_bu_a$ is the 4-acceleration vector (with $\sigma_{ab}u^b=0=\omega_{ab}u^b= \dot{u}_au^a$ by construction).\footnote{Round brackets denote symmetrisation and square ones antisymmetrisation. Angled brackets, on the other hand, indicate the symmetric and traceless part of spacelike tensors. For instance, ${\rm D}_{\langle a}u_{b\rangle}= {\rm D}_{(a}u_{b)}-({\rm D}_cu^c/3)h_{ab}$.} Positive values for $\Theta$ mean that the tangent worldlines expand and negative ones imply contraction. The shear describes distortions in the shape of the congruence under constant volume. Nonzero vorticity, on the other hand, indicates that the worldlines are rotating relative to each other. Finally, the 4-acceleration manifests the presence of non-gravitational forces. We also note that, on using the spatial Levi-Civita tensor, we may define the vorticity vector $\omega_a=\varepsilon_{abc} \omega^{bc}/2$. The latter determines the rotational axis.

\subsection{1+1+2 splitting}\label{ss1+1+2S}
Decomposing the 4-dimensional spacetime into time and 3-dimensional space may not be enough when the spatial sections are anisotropic. Suppose there is a preferred spatial direction and $n_a$ is the unit vector parallel to it. Then $u_an^a=0$ and $n_an^a=1$ by construction, while the tensor
\begin{equation}
\tilde{h}_{ab}= h_{ab}- n_an_b\,,  \label{thab1}
\end{equation}
projects into the 2-dimensional spacelike surfaces orthogonal to $n_a$. Indeed, following (\ref{thab1}), we obtain $\tilde{h}_{ab}u^b=0=\tilde{h}_{ab}n^b$, while one can easily verify that $\tilde{h}_{ab}\tilde{h}^b{}_c= \tilde{h}_{ac}$ and $\tilde{h}_a{}^a=2$. The $n_a$-field and the $\tilde{h}_{ab}$-tensor decompose the 3-dimensional space into a spatial direction parallel to $n_a$ and 2-dimensional spacelike surfaces (``sheets'') normal to $n_a$ (see~\cite{CB,C} for details). Therefore, we have achieved an overall 1+1+2 splitting of the spacetime into a temporal direction (along $u_a$), a spatial direction (parallel to $n_a$) and 2-dimensional spacelike surfaces orthogonal to both of these vectors. This decomposition is reflected in the following splitting
\begin{equation}
g_{ab}= \tilde{h}_{ab}+ n_an_b- u_au_b\,,  \label{thab2}
\end{equation}
of the spacetime metric.\footnote{The alternating Levi-Civita tensor of the 2-D surfaces orthogonal to the $n_a$-field (i.e.~the area element $\tilde{\varepsilon}_{ab}$) is defined as the contraction of its 3-D associate along $n_a$. In particular, we define $\tilde{\varepsilon}_{ab}=\varepsilon_{abc}n^c$ and $\varepsilon_{abc}=n_a\varepsilon_{bc}+n_b\varepsilon_{ca}+ n_c\varepsilon_{ab}$. Then, $\tilde{\varepsilon}_{ab}= \tilde{\varepsilon}_{[ab]}=\pm\varepsilon_{12}=\pm1$, with $\tilde{\varepsilon}_{ab}u^b=0= \tilde{\varepsilon}_{ab}n^b$ and $\tilde{\varepsilon}_{ab}\tilde{\varepsilon}^{cd}= 2\tilde{h}_{[a}{}^c\tilde{h}_{b]}{}^d$ by construction. The latter relation immediately leads to $\tilde{\varepsilon}_{ac} \tilde{\varepsilon}^{bc}= \tilde{h}_a{}^b$ and $\tilde{\varepsilon}_{ab}\tilde{\varepsilon}^{ab}=2$~\citep{CB,C}.} Moreover, in direct analogy with definitions (\ref{1+3dirs}a) and (\ref{1+3dirs}b), the  derivatives parallel and orthogonal to the $n_a$-field are defined by~\citep{CB,C}
\begin{equation}
T^{\,\prime}_{ab\cdots}{}^{cd\cdots}= n^s{\rm D}_sT_{ab\cdots}{}^{cd\cdots} \hspace{10mm} {\rm and} \hspace{10mm} \tilde{\rm D}_sT_{ab\cdots}{}^{cd\cdots}= \tilde{h}_s{}^q\tilde{h}_a{}^f\tilde{h}_b{}^k\cdots \tilde{h}_p{}^c\tilde{h}_r{}^d\cdots{\rm D}_qT_{fk\cdots}{}^{pr\cdots}\,.  \label{1+2dirs}
\end{equation}
Applying the operators (\ref{1+2dirs}a) and (\ref{1+2dirs}b) to the 2-D projector ($\tilde{h}_{ab}$), using definition (\ref{thab1}) and keeping in mind that ${\rm D}_ch_{ab}=0$, provides the auxiliary relations
\begin{equation}
\tilde{h}_{ab}^{\prime}= -2n_{(a}^{\prime}n_{b)} \hspace{10mm} {\rm and} \hspace{10mm} \tilde{\rm D}_c\tilde{h}_{ab}= 0\,,  \label{thdirs}
\end{equation}
respectively. The latter result implies that the $\tilde{h}_{ab}$ can act as the metric of the associated 2-surfaces, in the same way $h_{ab}$ can be seen as the metric of the spatial hypersurfaces. We also note that the vectors $u_a$ and $n_a$ are globally orthogonal to the corresponding 3-surfaces and 2-surfaces when they are irrotational. Otherwise their orthogonality is only local.

The ``kinematics'' of the $n_a$-field are monitored by a set of  irreducible variables, obtained in a manner exactly analogous to the one used for the 4-velocity vector (see \S~\ref{ss1+3S} before). More specifically, employing definitions (\ref{1+2dirs}a) and (\ref{1+2dirs}b), gives
\begin{equation}
{\rm D}_bn_a= {1\over2}\,\tilde{\Theta}\tilde{h}_{ab}+ \tilde{\sigma}_{ab}+ \tilde{\omega}_{ab}+ n^{\prime}_an_b\,,  \label{Dbna}
\end{equation}
with $\tilde{\Theta}=\tilde{\rm D}_an^a$, $\tilde{\sigma}_{ab}=\tilde{\rm D}_{\langle b}n_{a\rangle}$, $\tilde{\omega}_{ab}=\tilde{\rm D}_{[b}n_{a]}$ and $n^{\prime}_a=n^b{\rm D}_bn_a$. Note that $\tilde{\sigma}_{ab}n^b=0=\tilde{\omega}_{ab}n^b=n^{\prime}_an^a$ by construction.\footnote{The time-derivative of $n_a$ decomposes as $\dot{n}_a= \dot{u}^bn_bu_a+\tilde{h}_a{}^b\dot{n}_b$, where the first term on the right-hand side is purely temporal and the second is confined to the 2-dimensional sheet orthogonal to $n_a$~\citep{CB,C}.} The physical/geometrical interpretation of $\tilde{\Theta}$, $\tilde{\sigma}_{ab}$, $\tilde{\omega}_{ab}$ and $n^{\prime}_a$, is closely analogous to that of their 3-dimensional counterparts (see \S~\ref{ss1+3S} before). In particular, suppose that the $n_a$-field is tangent to a congruence of spacelike curves and consider a 2-dimensional cross-section ($S$) of this congruence. Then, positive/negative values of the area scalar $\tilde{\Theta}$ imply that the aforementioned curves converge/diverge. In other words, the congruence expands/contracts and the area of $S$ increases/decreases accordingly. The symmetric and trace-free 2-tensor $\tilde{\sigma}_{ab}$ is analogous to the shear tensor defined in the previous section and monitors changes in the shape of $S$, under constant area. On the other hand, the antisymmetric 2-tensor $\tilde{\omega}_{ab}$ describes the rotational behaviour of the congruence. Note that the antisymmetry of $\tilde{\omega}_{ab}$ means that the latter has only one independent component. We may therefore write $\tilde{\omega}_{ab}=\tilde{\omega}\tilde{\varepsilon}_{ab}$, with $\tilde{\varepsilon}_{ab}=\varepsilon_{abc}n^c$ representing the 2-dimensional Levi-Civita tensor and $\tilde{\omega}= \tilde{\omega}_{ab}\tilde{\varepsilon}^{ab}/2$. Finally, the 2-vector $n^{\prime}_a$ vanishes when the curves in question are spacelike geodesics.

\section{Kinematics of spacelike congruences}\label{sKSCs}
As with the timelike worldlines, the kinematics of spacelike congruences are determined by a set of propagation formulae, which describe the evolution of the associated area element ($\tilde{\Theta}$), surface shear ($\tilde{\sigma}_{ab}$) and surface vorticity ($\tilde{\omega}_{ab}$), along the direction of the congruence.

\subsection{Irreducible kinematic evolution}\label{ssIKE}
The kinematic evolution of a timelike congruence follows after applying the 4-dimensional Ricci identity to the corresponding 4-velocity field (e.g.~see \S~1.3.1 in~\cite{TCM}). In analogy, the kinematics of a spacelike vector-field follow from the Ricci identity of the spatial sections. Applied to an arbitrary spacelike vector $v_a$, the latter reads (e.g.~see Appendix~A.3 in~\cite{TCM})
\begin{equation}
2{\rm D}_{[a}{\rm D}_{b]}v_c=-2\omega_{ab}\dot{v}_c+ \mathcal{R}_{dcba}v^d\,,  \label{3Ricci}
\end{equation}
with $\mathcal{R}_{abcd}$ being the 3-dimensional Riemann curvature tensor. For zero vorticity the latter satisfies all the symmetries of its spacetime counterpart. Otherwise we have $\mathcal{R}_{abcd}= \mathcal{R}_{[ab][cd]}$ only (see \S~1.3.5 in~\cite{TCM} for details). Assuming that $v_a\equiv n_a$, where $n_a$ is a unit spacelike vector (i.e.~$u_an^a=0$ and $n_an^a=1$), contracting (\ref{3Ricci}) along $n_a$ and using decomposition (\ref{Dbna}), we obtain\footnote{In deriving the intermediate formula (\ref{Dbna'1}), we have also employed the auxiliary expression
\begin{equation}
{\rm D}_bn_a^{\prime}= \tilde{\rm D}_bn_a^{\prime}- {1\over2}\,\tilde{\Theta}n_an_b^{\prime}- n_a(\tilde{\sigma}_{bc}-\tilde{\omega}_{bc})n^{\prime\,c}+ (n_a^{\prime}n_b)^{\prime}- n_a^{\prime}n_b^{\prime}\,.  \label{aux1}
\end{equation}}
\begin{eqnarray}
\left({\rm D}_bn_a\right)^{\prime}&=& -{1\over4}\,\tilde{\Theta}^2\tilde{h}_{ab}- \tilde{\Theta}(\tilde{\sigma}_{ab}+\tilde{\omega}_{ab})- \tilde{\sigma}_{ca}\tilde{\sigma}_b{}^c- \tilde{\omega}_{ca}\tilde{\omega}_b{}^c+ 2\tilde{\sigma}_{c[a}\tilde{\omega}_{b]}{}^c\nonumber\\ &&+\tilde{\rm D}_bn^{\prime}_a- \tilde{\Theta}n_{(a}n^{\prime}_{b)}- 2n_{(a}\tilde{\sigma}_{b)c}n^{\prime\,c}+ 2n_{[a}\tilde{\omega}_{b]c}n^{\prime\,c}+ (n^{\prime}_an_b)^{\prime}- n^{\prime}_an^{\prime}_b\nonumber\\ &&-\mathcal{R}_{acbd}n^cn^d+ 2\dot{n}_a\omega_{bc}n^c\,. \label{Dbna'1}
\end{eqnarray}
Substituting (\ref{Dbna}) into the left-hand side of the above and recalling that $\tilde{h}_{ab}^{\prime}= -2n_{(a}n_{b)}^{\prime}$ (see Eq.~(\ref{thdirs}a) in \S~\ref{ss1+1+2S}), gives
\begin{eqnarray}
{1\over2}\,\tilde{\Theta}^{\prime}\tilde{h}_{ab}+ \tilde{\sigma}_{ab}^{\prime}+ \tilde{\omega}_{ab}^{\prime}&=& -{1\over4}\,\tilde{\Theta}^2\tilde{h}_{ab}- \tilde{\Theta}(\tilde{\sigma}_{ab}+\tilde{\omega}_{ab})- \tilde{\sigma}_{ca}\tilde{\sigma}_b{}^c- \tilde{\omega}_{ca}\tilde{\omega}_b{}^c+ 2\tilde{\sigma}_{c[a}\tilde{\omega}_{b]}{}^c\nonumber\\ &&+\tilde{\rm D}_bn^{\prime}_a- 2n_{(a}\tilde{\sigma}_{b)c}n^{\prime\,c}+ 2n_{[a}\tilde{\omega}_{b]c}n^{\prime\,c}- n^{\prime}_an^{\prime}_b\nonumber\\ &&-\mathcal{R}_{acbd}n^cn^d+ 2\dot{n}_a\omega_{bc}n^c\,.  \label{Dbna'2}
\end{eqnarray}
Finally, projecting orthogonal to $n_a$ and keeping in mind that $\mathcal{R}_{abcd}= \mathcal{R}_{[ab][cd]}$, we arrive at
\begin{eqnarray}
{1\over2}\,\tilde{\Theta}^{\prime}\tilde{h}_{ab}+ \tilde{h}_{\langle a}{}^c\tilde{h}_{b\rangle}{}^d \tilde{\sigma}_{cd}^{\prime}+ \tilde{h}_{[a}{}^c\tilde{h}_{b]}{}^d \tilde{\omega}_{cd}^{\prime}&=& -{1\over4}\,\tilde{\Theta}^2\tilde{h}_{ab}- \tilde{\Theta}(\tilde{\sigma}_{ab}+\tilde{\omega}_{ab})- \tilde{\sigma}_{ca}\tilde{\sigma}_b{}^c- \tilde{\omega}_{ca}\tilde{\omega}_b{}^c+ 2\tilde{\sigma}_{c[a}\tilde{\omega}_{b]}{}^c\nonumber\\ &&+\tilde{\rm D}_bn^{\prime}_a- n^{\prime}_an^{\prime}_b- \mathcal{R}_{acbd}n^cn^d+ 2\tilde{h}_a{}^c\dot{n}_c\omega_{bd}n^d\,,  \label{Dbna'3}
\end{eqnarray}
given that $\tilde{h}_a{}^c\tilde{h}_b{}^d \tilde{\sigma}_{cd}^{\prime}=\tilde{h}_{\langle a}{}^c \tilde{h}_{b\rangle}{}^d\tilde{\sigma}_{cd}^{\prime}$ and that $\tilde{h}_a{}^c\tilde{h}_b{}^d\tilde{\omega}_{cd}^{\prime}= \tilde{h}_{[a}{}^c\tilde{h}_{b]}{}^d \tilde{\omega}_{cd}^{\prime}$. This expression monitors the evolution of the spacelike congruence tangent to the unitary $n_a$-field, along the (spatial) direction of the latter. More specifically, the trace, the projected symmetric trace-free and the projected antisymmetric components of (\ref{Dbna'2}) provide the evolution formulae of the area scalar ($\tilde{\Theta}$), of the 2-shear tensor ($\tilde{\sigma}_{ab}$) and of the 2-vorticity  tensor ($\tilde{\omega}_{ab}$) respectively.

\subsection{Raychaudhuri's formula for spacelike
congruences}\label{ssRFSCs}
Taking the trace of (\ref{Dbna'3}), while keeping in mind that $\tilde{h}_{ab}n^b=0$ and $\tilde{\sigma}_{ab}n^b=0= \tilde{\omega}_{ab}n^b=n^{\prime}_an^a$, we obtain the following 3-dimensional analogue of the Raychaudhuri equation
\begin{equation}
\tilde{\Theta}^{\prime}= -{1\over2}\,\tilde{\Theta}^2- \mathcal{R}_{ab}n^an^b- 2\left(\tilde{\sigma}^2-\tilde{\omega}^2\right)+ \tilde{\rm D}_an^{\prime\,a}- n_a^{\prime}n^{\prime\,a}+ 2\omega_{ab}\dot{n}^an^b\,,  \label{3Ray1}
\end{equation}
which monitors the evolution of the area scalar $\tilde{\Theta}$ along the $n_a$-direction.\footnote{Comparing (\ref{3Ray1}) to the (standard) Raychaudhuri equation of a timelike congruence (e.g.~see expression (1.3.3) in~\cite{TCM}), one notices that only the last term on the right-hand side of (\ref{3Ray1}) has no 4-dimensional analogue. When the host spacetime is irrotational, the aforementioned extra term vanishes. In that case the only (formalistic) difference between Eqs.~(\ref{3Ray1}) here and (1.3.3) in~\cite{TCM}, is in the sign of the second-last term. This difference reflects the fact that $h_{ab}$ is orthogonal to a timelike vector field, whereas $\tilde{h}_{ab}$ is normal to a spacelike vector.} Note that $\mathcal{R}_{ab}=h^{cd}\mathcal{R}_{cadb}= \mathcal{R}^c{}_{acb}$ defines the 3-D Ricci tensor, which is not necessarily symmetric (see Eq.~(\ref{3Rab}) below). Also, $\tilde{\sigma}^2=\tilde{\sigma}_{ab}\tilde{\sigma}^{ab}/2$ and $\tilde{\omega}^2=\tilde{\omega}_{ab}\tilde{\omega}^{ab}/2$ by construction. As in the standard Raychaudhuri equation of timelike worldlines, positive terms on the right-hand side of the above force our spacelike congruence to diverge, while negative ones lead to its convergence.

When dealing with a congruence of spacelike geodesics (i.e.~where $n_a^{\prime}=n^b{\rm D}_bn_a=0$ by default), expression (\ref{3Ray1}) reduces to
\begin{equation}
\tilde{\Theta}^{\prime}= -{1\over2}\,\tilde{\Theta}^2- \mathcal{R}_{ab}n^an^b- 2\left(\tilde{\sigma}^2-\tilde{\omega}^2\right)+ 2\omega_{ab}\dot{n}^an^b\,.  \label{3Ray2}
\end{equation}
Moreover, when the host spacetime is not rotating, the $u_a$-field is also irrotational (i.e.~$\omega_{ab}=0$) and the last term of above vanishes identically. In that case, the antisymmetric component of the 3-Ricci tensor vanishes as well (i.e.~$\mathcal{R}_{ab}=\mathcal{R}_{(ab)}$ -- see Eq.~(\ref{3Rab}) next).

From the purely gravitational point of view, the key variable on the right-hand side of Eqs.~(\ref{3Ray1}) and (\ref{3Ray2}) is the 3-Ricci tensor. The latter determines the curvature of the 3-D hypersurfaces orthogonal to $u_a$ and also carries the effect of the matter fields. Following~\cite{TCM}, we note that (unlike its 4-dimensional counterpart) $\mathcal{R}_{ab}$ is not necessarily symmetric and it is given by
\begin{eqnarray}
\mathcal{R}_{ab}&=& {2\over3} \left(\kappa\rho-{1\over3}\,\Theta^2 +\sigma^2-\omega^2\right)h_{ab}+ E_{ab}+ {1\over2}\,\kappa\pi_{ab}- {1\over3}\,\Theta(\sigma_{ab}+\omega_{ab})+ \sigma_{c\langle a}\sigma_{b\rangle}{}^c\nonumber\\ &&+\omega_{c\langle a}\omega_{b\rangle}{}^c- 2\sigma_{c[a}\omega_{b]}{}^c\,,  \label{3Rab}
\end{eqnarray}
where $\kappa=8\pi G$ is the gravitational constant. Here, $\Theta$, $\sigma_{ab}$ and $\omega_{ab}$ are the irreducible kinematic variables of the $u_a$-field (see \S~\ref{ss1+3S} earlier), with $\sigma^2=\sigma_{ab}\sigma^{ab}/2$ and $\omega^2=\omega_{ab}\omega^{ab}/2$. Also, $\rho$ and $\pi_{ab}$ are respectively the energy density and the anisotropic pressure of the total matter, while $E_{ab}$ is the electric part of the Weyl tensor (all measured relative to the $u_a$-field). The Weyl field monitors the action of gravity at a distance, namely tidal forces and gravitational waves. Finally, we note that $\pi_{ab}=\pi_{\langle ab\rangle}$, $E_{ab}=E_{\langle ab\rangle}$ and $\pi_{ab}u^b=0=E_{ab}u^b$ (e.g.~see \S~1.3.5 in~\cite{TCM}).

\subsection{Shear and vorticity evolution}\label{ssSVE}
The symmetric trace-free and the antisymmetric parts of (\ref{Dbna'3}) govern the evolution of the 2-shear and the 2-vorticity tensors, along the direction of $n_a$. More specifically, we obtain
\begin{eqnarray}
\tilde{h}_{\langle a}{}^c\tilde{h}_{b\rangle}{}^d \tilde{\sigma}_{cd}^{\prime}&=& -\tilde{\Theta}\tilde{\sigma}_{ab}- \tilde{\sigma}_{c\langle a}\tilde{\sigma}_{b\rangle}{}^c- \tilde{\omega}_{c\langle a}\tilde{\omega}_{b\rangle}{}^c+ \tilde{\rm D}_{\langle b}n^{\prime}_{a\rangle}- n^{\prime}_{\langle a}n^{\prime}_{b\rangle}- \mathcal{R}_{\langle a}{}^c{}_{b\rangle}{}^dn_cn_d\nonumber\\ &&+2\tilde{h}_{c\langle a}\omega_{b\rangle d}\dot{n}^cn^d \label{2sh'1}
\end{eqnarray}
and
\begin{equation}
\tilde{h}_{[a}{}^c\tilde{h}_{b]}{}^d \tilde{\omega}_{cd}^{\prime}= -\tilde{\Theta}\tilde{\omega}_{ab}+ 2\tilde{\sigma}_{c[a}\tilde{\omega}_{b]}{}^c+ \tilde{\rm D}_{[b}n^{\prime}_{a]}- \mathcal{R}_{[a}{}^c{}_{b]}{}^dn_cn_d+ 2\tilde{h}_{c[a}\omega_{b]d}\dot{n}^cn^d\,,  \label{2vort'1}
\end{equation}
for the 2-shear and the 2-vorticity tensors respectively. When the $n_a$-congruence is geodesic and the 4-velocity field is irrotational (i.e.~for $n_a^{\prime}=0= \omega_{ab}$), the above two expressions simplify to
\begin{equation}
\tilde{\sigma}_{ab}^{\prime}= -\tilde{\Theta}\tilde{\sigma}_{ab}- \tilde{\sigma}_{c\langle a}\tilde{\sigma}_{b\rangle}{}^c- \tilde{\omega}_{c\langle a}\tilde{\omega}_{b\rangle}{}^c- \mathcal{R}_{\langle a}{}^c{}_{b\rangle}{}^dn_cn_d  \label{2sh'2}
\end{equation}
and
\begin{equation}
\tilde{\omega}_{ab}^{\prime}= -\tilde{\Theta}\tilde{\omega}_{ab}+ 2\tilde{\sigma}_{c[a}\tilde{\omega}_{b]}{}^c- \mathcal{R}_{[a}{}^c{}_{b]}{}^dn_cn_d\,,  \label{2vort'2}
\end{equation}
respectively. Therefore, vorticity sources shear but the opposite is not necessarily true. Also, spatial curvature generally affects the evolution of both $\tilde{\sigma}_{ab}$ and $\tilde{\omega}_{ab}$.

As with the Raychaudhuri equation before, the effect of the matter fields is carried by the curvature terms. In a general spacetime, the Riemann tensor of the 3-dimensional hypersurfaces is given by the expression (see~\S~1.3.5 in~\cite{TCM})
\begin{eqnarray}
\mathcal{R}_{abcd}&=& -\varepsilon_{abq}\varepsilon_{cds}E^{qs}+ {1\over3}\left(\kappa\rho-{1\over3}\,\Theta^2\right) (h_{ac}h_{bd}-h_{ad}h_{bc})\nonumber\\ &&+{1\over2}\,\kappa\,(h_{ac}\pi_{bd}+\pi_{ac}h_{bd} -h_{ad}\pi_{bc}-\pi_{ad}h_{bc})\nonumber\\ &&-{1\over3}\,\Theta[h_{ac}(\sigma_{bd}+\omega_{bd}) +(\sigma_{ac}+\omega_{ac})h_{bd} -h_{ad}(\sigma_{bc}+\omega_{bc}) -(\sigma_{ad}+\omega_{ad})h_{bc}]\nonumber\\  &&-(\sigma_{ac}+\omega_{ac})(\sigma_{bd}+\omega_{bd}) +(\sigma_{ad}+\omega_{ad})(\sigma_{bc}+\omega_{bc})\,,  \label{3Riemann}
\end{eqnarray}
guaranteeing that $\mathcal{R}_{abcd}= \mathcal{R}_{[ab][cd]}$ always and that $\mathcal{R}_{abcd}=\mathcal{R}_{cdab}$ only when $\omega_{ab}=0$. Substituting the above into the right-hand side of (\ref{2sh'1}) and (\ref{2vort'1}) leads to
\begin{eqnarray}
\tilde{h}_{\langle a}{}^c\tilde{h}_{b\rangle}{}^d \tilde{\sigma}_{cd}^{\prime}&=& -\tilde{\Theta}\tilde{\sigma}_{ab}- \tilde{\sigma}_{c\langle a}\tilde{\sigma}_{b\rangle}{}^c- \tilde{\omega}_{c\langle a}\tilde{\omega}_{b\rangle}{}^c+ \tilde{\rm D}_{\langle b}n^{\prime}_{a\rangle}- n^{\prime}_{\langle a}n^{\prime}_{b\rangle}+\varepsilon_{\langle a}{}^{cq}\varepsilon_{b\rangle}{}^{ds}n_cn_dE_{qs}\nonumber\\ &&+ {1\over3}\left(\kappa\rho-{1\over3}\,\Theta^2\right)n_{\langle a}n_{b\rangle}- {1\over2}\,\kappa\left(\pi_{ab}-2n_{\langle a} \pi_{b\rangle}{}^cn_c\right)+ {1\over3}\,\Theta\left(\sigma_{ab}- 2n_{\langle a}\sigma_{b\rangle}{}^cn_c\right)\nonumber\\ &&+\sigma_{ab}\sigma^{cd}n_cn_d- \sigma_{\langle a}{}^c\sigma_{b\rangle}{}^dn_cn_d+ \omega_{\langle a}{}^c\omega_{b\rangle}{}^dn_cn_d+ 2\tilde{h}_{c\langle a}\omega_{b\rangle d}\dot{n}^cn^d \label{2sh'3}
\end{eqnarray}
and
\begin{eqnarray}
\tilde{h}_{[a}{}^c\tilde{h}_{b]}{}^d \tilde{\omega}_{cd}^{\prime}&=& -\tilde{\Theta}\tilde{\omega}_{ab}+ 2\tilde{\sigma}_{c[a}\tilde{\omega}_{b]}{}^c+ \tilde{\rm D}_{[b}n^{\prime}_{a]}+ {1\over3}\,\Theta\left(\omega_{ab}+2n_{[a}\omega_{b]}{}^cn_c\right)+ \omega_{ab}\sigma^{cd}n_cn_d\nonumber\\ &&-2\omega_{[a}{}^c\sigma_{b]}{}^dn_cn_d+ 2\tilde{h}_{c[a}\omega_{b]d}\dot{n}^cn^d\,,  \label{2vort'3}
\end{eqnarray}
respectively. Note the absence of any geometric or matter terms in the latter expression. This shows that the geometry of the host spacetime, namely the gravitational field, does not affect (at least directly) the rotational behaviour of spacelike congruences. According to Eq.~(\ref{2sh'3}), on the other hand, this is not the case for shear-like deformations.

Before closing this section we should emphasise that the formulae derived so far are purely geometrical in nature and depend solely on the structure of the 3-dimensional hypersurfaces and on that of their host spacetime. Also, no specific assumptions have been made about the material content, the effects of which enter into the equations through the 3-Riemann and the 3-Ricci tensors.

\section{The magnetic-field case}\label{sMFC}
Magnetism is an integrable part of the cosmos with a verified presence almost everywhere in the universe. Also, magnetic fields are rather unique matter sources and what distinguishes them from the rest is their vector nature and tension properties. In what follows we will use the formalism developed so far to look closer into the implications of these special magnetic features.

\subsection{Magnetic pressure and magnetic tension}\label{ssMPMT}
Consider the 4-dimensional spacetime defined in \S~\ref{sSD} earlier. Relative to observers moving with a timelike 4-velocity $u_a$, the electromagnetic tensor ($F_{ab}=F_{[ab]}$) decomposes into its electric and magnetic parts. These are respectively given by~\citep{TCM}
\begin{equation}
E_a= F_{ab}u^b \hspace{10mm} {\rm and} \hspace{10mm} B_a= {1\over2}\,\varepsilon_{abc}F^{bc}\,,  \label{EaBa}
\end{equation}
with $\varepsilon_{abc}$ being the 3-dimensional Levi-Civita tensor (see footnote~2 earlier). Then, $E_au^a=0=B_au^a$, to guarantee that both the electric and the magnetic fields are spacelike vectors.

Let us concentrate on the magnetic component of the Maxwell field and switch its electric counterpart off, as it happens in the ideal magnetohydrodynamic (MHD) limit for example. In such a case, the electromagnetic stress-energy tensor reduces to
\begin{equation}
T_{ab}= \rho_{_B}u_au_b+ p_{_B}h_{ab}+ \Pi_{ab}\,,  \label{MHDTab}
\end{equation}
where $\rho_{_B}=B^2/2$ is the energy density, $p_{_B}=B^2/6$ is the isotropic pressure and $\Pi_{ab}=\Pi_{\langle ab\rangle}= (B^2/3)h_{ab}-B_aB_b$ is the anisotropic pressure of the $B$-field (with $B^2=B_aB^a$).\footnote{We use natural units for the matter and Heaviside-Lorentz units for the electromagnetic field.} The symmetric and trace-free $\Pi_{ab}$-tensor also carries the tension properties of the magnetic forcelines. The magnetic tension reflects the elasticity of the field-lines and their tendency to remain ``straight''. On the other hand, the total pressure exerted by the $B$-field (isotropic plus anisotropic) is encoded in the symmetric Maxwell tensor $\mathcal{M}_{ab}= (B^2/2)h_{ab}-B_aB_b$ (e.g.~see~\cite{Pa,Me}).

Suppose now that $\ell_a$ and $k_a$ are unitary spacelike vectors orthogonal and parallel to the magnetic field respectively. Then, $\ell_au^a=0=k_au^a$, with $\ell_aB^a=0$ and $B_a=Bk_a$ (where $B=\sqrt{B_aB^a}$). It is straightforward to show that both $\ell_a$ and $k_a$ are eigenvectors of the Maxwell tensor, though their associated eigenvalues have opposite signs. Indeed, projecting $\mathcal{M}_{ab}$ along $\ell_a$ gives a positive eigenvalue (i.e.~$\mathcal{M}_{ab}\ell^b=(1/2)\ell_a$), thus ensuring a (positive) magnetic pressure orthogonal to the field lines. Projecting along $k_a$, on the other hand, leads to a negative eigenvalue ($\mathcal{M}_{ab}k^b=-(1/2)k_a$), which implies that the $B$-field exerts a negative pressure (i.e.~a tension) along its own direction. Physically speaking, the magnetic pressure reflects the tendency of the forcelines to push each other apart, while the field's tension manifests the elasticity of the field lines, namely their tendency to remain ``straight'' and to react against any agent that distorts them from equilibrium~\citep{Pa,Me}.

\subsection{Magnetic-line convergence and focusing}\label{ssM-LCF}
Let us introduce a congruence of magnetic lines tangent to the field vector. Suppose also that $k_a$, with $B_a=Bk_a$, is the unitary spacelike vector along the direction of the the $B$-field (see \S~\ref{ssMPMT} above). Like any other source of energy, the magnetic field contributes to the total gravitational field through its energy density, pressure and tension (see Eq.~(\ref{MHDTab}) in \S~\ref{ssMPMT}). The question we would like to address is how gravity affects the convergence/divergence of the magnetic forcelines and, more specifically, whether the $B$-field will collapse under its own gravitational pull or not.

A family of spacelike curves will converge and focus when their 2-dimensional cross-sectional area becomes progressively smaller (along their own direction). In the opposite case the aforementioned congruence will diverge. Assuming that $n_a$ is the unit vector tangent to the aforementioned lines, changes in the size of their cross section are monitored by the divergence $\tilde{\Theta}=\tilde{\rm D}_an^a$, as defined in \S~\ref{ss1+1+2S} earlier. The evolution of $\tilde{\Theta}$ in the direction of the lines, namely along $n_a$, follows from the associated Raychaudhuri formula (see Eq.~(\ref{3Ray1}) in \S~\ref{ssRFSCs}). When dealing with the forcelines of a magnetic field, that is when $n_a\equiv k_a$, the latter reads
\begin{equation}
\tilde{\Theta}^{\prime}= -{1\over2}\,\tilde{\Theta}^2- \mathcal{R}_{ab}k^ak^b- 2\left(\tilde{\sigma}^2-\tilde{\omega}^2\right)+ \tilde{\rm D}_ak^{\prime\,a}- k_a^{\prime}k^{\prime\,a}+ 2\omega_{ab}\dot{k}^ak^b\,,  \label{M3Ray1}
\end{equation}
where $\mathcal{R}_{ab}$ is given by (\ref{3Rab}).\footnote{We remind the reader that the Raychaudhuri formula given in Eq.~(\ref{M3Ray1}) monitors the convergence/divergence, of the (spacelike) magnetic forcelines along their own (spatial) direction. Therefore, one should not confuse expression (\ref{M3Ray1}) with the Raychaudhuri equation monitoring the (timelike) worldlines of charged particles and their temporal evolution in the presence of a magnetic field (e.g.~see~\cite{R2,KT}).} Projecting the latter along the direction of the magnetic forcelines, while assuming the presence of other matter sources (with total energy density $\rho$ and anisotropic pressure $\pi_{ab}$), we obtain
\begin{eqnarray}
\mathcal{R}_{ab}k^ak^b&=& {2\over3} \left[\kappa\left(\rho+\rho_{_B}\right)-{1\over3}\,\Theta^2+\sigma^2-\omega^2\right]+ E_{ab}k^ak^b+ {1\over2}\,\kappa\left(\pi_{ab}+\Pi_{ab}\right)k^ak^b- {1\over3}\,\Theta\sigma_{ab}k^ak^b\nonumber\\ &&+\sigma_{c\langle a}\sigma_{b\rangle}{}^ck^ak^b+ \omega_{c\langle a}\omega_{b\rangle}{}^ck^ak^b\,,  \label{M3Rab2}
\end{eqnarray}
given that $h_{ab}k^ak^b=k_ak^a=1$. However, given that $\rho_{_B}=B^2/2$ and that $\Pi_{ab}=(1/3)B^2h_{ab}-B_aB_b$, we find that $(2/3)\kappa\rho_{_B}+(1/2)\kappa\Pi_{ab}k^ak^b=0$. Then,
\begin{eqnarray}
\mathcal{R}_{ab}k^ak^b&=& {2\over3} \left(\kappa\rho-{1\over3}\,\Theta^2+\sigma^2-\omega^2\right)+ E_{ab}k^ak^b+ {1\over2}\,\kappa\pi_{ab}k^ak^b- {1\over3}\,\Theta\sigma_{ab}k^ak^b\nonumber\\ &&+\sigma_{c\langle a}\sigma_{b\rangle}{}^ck^ak^b+ \omega_{c\langle a}\omega_{b\rangle}{}^ck^ak^b\,.  \label{M3Rab3}
\end{eqnarray}
This ensures that the magnetic energy-density and pressure do not contribute to the right-hand side of Eq.~(\ref{M3Ray1}). In other words, although the convergence/divergence of the magnetic forcelines is directly affected by the gravitational pull of the other matter sources, it proceeds unaffected by the $B$-field's own gravity (i.e.~by the magnetic gravitational energy). The reason behind this counterintuitive behaviour is the magnetic tension, which cancels out the field's energy-density input to the right-hand side of (\ref{M3Rab2}), (\ref{M3Rab3}) and therefore to Eq.~(\ref{M3Ray1}) itself.

The above refer to a general congruence of magnetic forcelines in a general spacetime filled with other forms of matter, in addition to the $B$-field. Further physical insight on the role of the magnetic tension can be obtained by considering the idealised case of forcelines that are irrotational and shear-free (spacelike) geodesics, resting in an otherwise empty and static space. Then, (\ref{M3Ray1}) reduces to
\begin{equation}
\tilde{\Theta}^{\prime}+ {1\over2}\,\tilde{\Theta}^2= -\mathcal{R}_{ab}k^ak^b\,,  \label{M3Ray2}
\end{equation}
with
\begin{equation}
\mathcal{R}_{ab}= {2\over3}\,\kappa\rho_{_B}h_{ab}+ {1\over2}\,\kappa\Pi_{ab}\,.  \label{M3Rab1}
\end{equation}
Keeping in mind that $\rho_{_B}=B^2/2$ and that $\Pi_{ab}=(1/3)B^2h_{ab}-B_aB_b$, the latter of the above gives $\mathcal{R}_{ab}k^ak^b=0$, which substituted back into Eq.~(\ref{M3Ray2}) leads to
\begin{equation}
\tilde{\Theta}^{\prime}= -{1\over2}\,\tilde{\Theta}^2\,,  \label{M3Ray3}
\end{equation}
ensuring that $\tilde{\Theta}^{\prime}=0$ at all times when $\tilde{\Theta}=0$ initially. This differential equation integrates immediately giving
\begin{equation}
\tilde{\Theta}= \tilde{\Theta}(\lambda)= {2\tilde{\Theta}_0\over2+\tilde{\Theta}_0\lambda}\,,  \label{tTheta}
\end{equation}
where $\tilde{\Theta}_0=\tilde{\Theta}(\lambda=0)$ and $\lambda$ may be seen as the proper length measured along the magnetic field lines. Accordingly, we may distinguish between the following three alternatives:
\begin{itemize}
\item When $\tilde{\Theta}_0<0$, we find that $\tilde{\Theta}\rightarrow-\infty$ as $\lambda\rightarrow-2/\tilde{\Theta}_0$
\item When $\tilde{\Theta}_0=0$, we have $\tilde{\Theta}=0$ at all times
\item When $\tilde{\Theta}_0>0$, we have $\tilde{\Theta}>0$ always
\end{itemize}
In other words, magnetic forcelines that are initially converging will focus to form caustics within finite proper length. If the lines happen to be stationary, on the other hand, they will remain so and will never converge. Finally, magnetic lines that are initially diverging will continue to do so indefinitely (since $\tilde{\Theta}\rightarrow0$ as $\lambda\rightarrow+\infty$ when $\tilde{\Theta}_0>0$ -- see solution (\ref{tTheta})). Note that (unlike typical timelike worldlines) in the last two cases the forcelines remain stationary, or keep diverging, despite the fact that the host 3-space is positively curved.\footnote{The mean curvature of the 3-space is decided by the trace of $\mathcal{R}_{ab}$. Recalling that $\rho_{_B}=B^2/2$ and that $\Pi_a{}^a=0$, we obtain $\mathcal{R}=\mathcal{R}_a{}^a=B^2$ to guarantee that the mean 3-curvature is positive (solely due to the magnetic presence).} Hence, although the spatial sections have positive mean curvature, the magnetic tension ensures that field lines will not ``feel'' the pull of their own gravity and therefore their ``motion'' is fully dictated by their initial condition.

The behaviour of the magnetic forcelines described so far is rather atypical and (to the best of our knowledge) particular to the $B$-field only. Indeed, consider the (spacelike) flow-lines of ordinary matter and assume that $t_a$ is their unit tangent vector. Assuming, for simplicity and demonstration purposes, that these lines are irrotational and shear-free geodesics, residing in an otherwise empty and static spacetime, the associated Raychaudhuri equation reads
\begin{equation}
\tilde{\Theta}^{\prime}+ {1\over2}\,\tilde{\Theta}^2= -\mathcal{R}_{ab}t^at^b\,,  \label{F3Ray1}
\end{equation}
with
\begin{equation}
\mathcal{R}_{ab}t^at^b= {2\over3}\,\kappa\rho+ {1\over2}\,\kappa\pi_{ab}t^at^b\,,  \label{F3Rab1}
\end{equation}
since $h_{ab}t^at^b=t_at^a=1$. In the case of a perfect fluid, with positive energy density ($\rho>0$) and zero viscosity ($\pi_{ab}=0$), we find that $\mathcal{R}_{ab}t^at^b>0$. Therefore, flow-lines that are initially static will converge and eventually focus (within finite length) under the pull of their own gravity alone. Also, in contrast to the $B$-field lines (see alternative No~3 above), initially diverging flow lines are not guaranteed to keep diverging. When dealing with an imperfect medium, however, the convergence of the flow-lines is not guaranteed, but depends on the sign and the magnitude of the anisotropic-pressure term ($\pi_{ab}t^at^b$) on the right-hand side of Eq.~(\ref{F3Rab1}). In particular, for matter with $\pi_{ab}t^at^b>-4\rho/3$ the flow-lines will definitely converge, but when $\pi_{ab}t^at^b< -4\rho/3$ the flow-lines may instead diverge. It is only for media with a magnetic-like ``equation of state'' (i.e.~for $\pi_{ab}t^at^b=-4\rho/3$) that the right-hand side of Eq.~(\ref{F3Ray1}) vanishes identically.

\subsection{Magnetic-line rotation and distortion}\label{ssM-LRD}
Following the evolution formula of the 2-vorticity (see Eq.~(\ref{2vort'3}) in \S~\ref{ssSVE}), the geometry and the matter content of the host spacetime do not affect the rotation of spacelike congruences. Hence, the rotation of the magnetic forcelines is not directly affected by the active gravitational field, including their own. Let us now turn to the 2-shear and apply expression (\ref{2sh'3}) to a set of magnetic forcelines residing in a general spacetime. Then, the 2-shear evolution formula reads
\begin{eqnarray}
\tilde{h}_{\langle a}{}^c\tilde{h}_{b\rangle}{}^d \tilde{\sigma}_{cd}^{\prime}&=& -\tilde{\Theta}\tilde{\sigma}_{ab}- \tilde{\sigma}_{c\langle a}\tilde{\sigma}_{b\rangle}{}^c- \tilde{\omega}_{c\langle a}\tilde{\omega}_{b\rangle}{}^c+ \tilde{\rm D}_{\langle b}n^{\prime}_{a\rangle}- n^{\prime}_{\langle a}n^{\prime}_{b\rangle}+\varepsilon_{\langle a}{}^{cq}\varepsilon_{b\rangle}{}^{ds}n_cn_dE_{qs}\nonumber\\ &&+ {1\over3}\left[\kappa\left(\rho+\rho_{_B}\right) -{1\over3}\,\Theta^2\right]n_{\langle a}n_{b\rangle}- {1\over2}\,\kappa\left(\pi_{ab}-2n_{\langle a}\pi_{b\rangle}{}^cn_c\right)- {1\over2}\,\kappa\left(\Pi_{ab}-2n_{\langle a}\Pi_{b\rangle}{}^cn_c\right)\nonumber\\ &&+{1\over3}\,\Theta\left(\sigma_{ab}- 2n_{\langle a}\sigma_{b\rangle}{}^cn_c\right)+ \sigma_{ab}\sigma^{cd}n_cn_d- \sigma_{\langle a}{}^c\sigma_{b\rangle}{}^dn_cn_d+ \omega_{\langle a}{}^c\omega_{b\rangle}{}^dn_cn_d\nonumber\\ &&+2\tilde{h}^c{}_{\langle a}\omega_{b\rangle d}\dot{n}^cn^d\,,  \label{M2sh'1}
\end{eqnarray}
where $\rho$, $\pi_{ab}$ and $\rho_{_B}$, $\Pi_{ab}$ are the energy density and the anisotropic pressure of the matter and of the $B$-field respectively. Then, given that $\rho_{_B}=B^2/2$ and that $\Pi_{ab}=(1/3)B^2h_{ab}-B_aB_b$, it is straightforward to show that the above reduces to Eq.~(\ref{2sh'3}) of \S~\ref{ssSVE}, with no explicit magnetic terms on the right-hand side ($\rho$ and $\pi_{ab}$ refer to the rest of the matter sources). Again, the absence of any direct magnetic effect is due to the field's tension, which cancels out the positive contribution from the magnetic energy density and pressure to Eq.~(\ref{M2sh'1}).

In summary, the convergence/divergence of the magnetic forcelines, their shear deformation and their rotation proceed unaffected by the $B$-field's own gravitational energy. Although the null effect on rotation applies to all spacelike congruences, the rest are entirely due to the field's tension. The latter guarantees that, although the magnetic lines of force respond to the gravitational pull of the other sources, they do not ``feel'' (at least not directly) their own gravity. This generic magnetic feature implies that (in the absence of other sources) a configuration of field lines that happens to be in equilibrium initially, will remain so indefinitely (unless an external agent interferes).

\section{Discussion}\label{sD}
Magnetic fields are ubiquitous and of rather unique nature, and what distinguishes them from the other known energy sources is their vector status and tension properties. In this work we have attempted to investigate the implications of the aforementioned features by looking into the ``kinematics'' of a congruence of magnetic forcelines. We did so by introducing an 1+2 splitting of the 3-dimensional space into a direction parallel to the field lines and 2-dimensional surfaces orthogonal to them. Taking a cross-sectional area of these lines, we defined three variables that monitor the area's expansion/contraction, rotation and shear-deformation. We then derived the equations describing the evolution of these variables along the direction of the magnetic lines of force. Our results showed that, although the magnetic congruence responds to the gravitational pull of the other sources, it is ``immune'' to its own gravity, no matter how strong the latter may be. More specifically, the kinematics of the magnetic forcelines are unaffected by the field's own contribution to the total gravitational energy. To the best of our knowledge, no other known matter source shows such a counterintuitive behaviour. The reason behind this unique magnetic conduct is its tension, which always cancels out the input of the field's energy density and isotropic pressure. In a sense, the magnetic tension ensures that the field lines do not ``feel'' their own gravitational pull. This also implies that, in a static and otherwise empty spacetime, a set of parallel magnetic forcelines will not converge or diverge, it will not rotate and it will not deform. Instead, the aforementioned congruence will remain in equilibrium until an external agent interferes.

These results are reminiscent of work done several decades ago, in the mid 1960s, by Melvin and Thorne~\citep{Mel1,Mel2,Th1,Th2}. It was shown, in particular, that there exists a stable solution of the Einstein-Maxwell equations that describes a cylindrical configuration of parallel magnetic forcelines in equilibrium, residing in an otherwise empty and static spacetime (as in our case -- see \S~\ref{ssM-LCF}, \S~\ref{ssM-LRD} above). This solution is also known as ``Melvin's magnetic universe''. It was also argued that \textit{``a pure magnetic field has a remarkable and previously unsuspected ability to stabilise itself against gravitational collapse''}.\footnote{It was claimed in~\cite{Mel1,Mel2} that pure electric fields show the same behaviour. The symmetry of the source-free Maxwell's equations guarantees that our results also apply to a pure electric field.} Whether this ability would be enough to avoid the ultimate singularity was left unanswered, but a number of crucial questions regarding the magnetic role during gravitational collapse was raised~\citep{Mel1,Mel2,Th1,Th2}. Our work seems to indicate that the magnetic tension, namely the elasticity of the field lines, may be the physical reason behind such a remarkable ability. This suggestion is also corroborated by other studies showing how the field's tension gives rise to ever increasing magneto-curvature stresses that resist the gravitational collapse of a magnetised medium~\citep{Ts1,Ts2,Ts3}. As with the work of~\cite{Mel1,Mel2,Th1,Th2}, however, the complexity of the problem made it impossible to establish whether such stresses would be capable of preventing the singularity from forming. Here, by treating the field lines as a congruence of spacelike curves, we have initiated a rather novel approach, which (once again) brought to the fore the role of the field's tension as a stabilising agent. Future work will try to exploit the advantages of such a treatment and shed more light on the potential magnetic implications for gravitational collapse


\begin{thebibliography}{}
\bibitem[Abreu \& Visser (2011)]{AV} Abreu G. and Visser M., Phys. Rev. D \textbf{83}, 10416 (2011).
\bibitem[Clarkson \& Barrett (2003)]{CB} Clarkson C.A. and Barrett R.K., Class. Quantum Grav. \textbf{20}, 3855 (2003).
\bibitem[Clarkson (2007)]{C} Clarkson C.A., Phys. Rev.D \textbf{76}, 104034 (2007).
\bibitem[Dadhich (2005)]{D1} Dadhich N., [arXiv: gr-qc/0511123]].
\bibitem[Dadhich (2007)]{D2} Dadhich N., Pramana J. Phys. \textbf{69}, 23 (2007).
\bibitem[Das (2014)]{Da} Das S., Phys. Rev. D \textbf{89}, 084068 (2014).
\bibitem[Dasgupta et al (2008)]{DNK1} Dasgupta A., Nandan H and Kar S., Ann. Phys. \textbf{323}, 1621 (2008).
\bibitem[Dasgupta et al (2009)]{DNK2} Dasgupta A., Nandan H and Kar S., Phys. Rev, D \textbf{79}, 124004 (2009).
\bibitem[Ehlers (2007)]{Eh} Ehlers J., Pramana J. Phys. \textbf{69}, 7 (2007).
\bibitem[Ellis (1971)]{El1} Ellis G.F.R., in \textit{General Relativity and Cosmology, Varenna Lectures}, Edit. R.K. Sachs (Academic Press, New York, 1971)~p.~104.
\bibitem[Ellis (1973)]{El2} Ellis G.F.R., in \textit{Garg\`ese Lectures in Physics}, Edit. E. Schatzman (Gordon and Breach, New York, 1973)~p.~1.
\bibitem[Ellis (2007)]{El3} Ellis G.F.R., Pramana J. Phys. \textbf{69}, 15 (2007).
\bibitem[Gannouji et al (2011)]{GND} Gannouji R., Nandan H. and Dadhich N., JCAP \textbf{11}, 051 (2011).
\bibitem[Greenberg (1970)]{G} Greenberg P.J., J. Math. Anal. Applic. \textbf{30}, 128 (1970).
\bibitem[Harko \& Lobo (2012)]{HL} Harko T. and Lobo F.S.N, Phys. Rev. D \textbf{86}, 124036 (2012).
\bibitem[Kantowski (1967)]{K} Kantowski R., J. Math. Phys. \textbf{9}, 336 (1967).
\bibitem[Kar \& Sengupta (2007)]{KS} Kar S. and Sengupta S., Pramana J. Phys. \textbf{69}, 23 (2007).
\bibitem[Kouretsis \& Tsagas (2010)]{KT} Kouretsis A.P. and Tsagas C.G., Phys. Rev. D \textbf{82}, 124053 (2010).
\bibitem[Kuniyal et al (2015)]{KUNZ} Kuniyal R.S., Uniyal R., Nandan H. and Zaidi A., Astrophys. Space Sci. \textbf{375}, 92 (2015).
\bibitem[Mason \& Tsamparlis (1985)]{MT} Mason D.P. and Tsamparlis M., J. Math. Phys. \textbf{26}, 2881 (1985).
\bibitem[Melvin (1964)]{Mel1} Melvin M.A., Phys. Lett. \textbf{8}, 65 (1964).
\bibitem[Melvin (1965)]{Mel2} Melvin M.A., Phys. Rev. B \textbf{139}, 225 (1965).
\bibitem[Mestel (2012)]{Me} Mestel L., \textit{Stellar Magnetism} (Oxford University Press, Oxford, 2012).
\bibitem[Mosheni (2015)]{Mo} Mohseni M., Gen. Rel. Grav. \textbf{47}, 24 (2015).
\bibitem[Pahwa et al (2015)]{PNG} Pahwa I., Nandan H. and Goswami U.D., Astrophys. Space Sci. \textbf{360}, 26 (2015).
\bibitem[Parker (1979)]{Pa} Parker E.N., \textit{Cosmical Magnetic Fields} (Oxford University Press, Oxford, 1979).
\bibitem[Pasmatsiou et al (2017)]{PTB} Pasmatsiou K., Tsagas C.G. and Barrow J.D., Phys. Rev. D \textbf{95}, 104007 (2017).
\bibitem[Poisson (2004)]{Po} Poisson E., \textit{A Relativist's Toolkit} (Cambridge University Press, Cambridge, 2004).
\bibitem[Raychaudhuri (1955)]{R1} Raychaudhuri A.K., Phys. Rev. \textbf{98}, 1123 (1955).
\bibitem[Raychaudhuri (1975)]{R2} Raychaudhuri A.K., Ann. Inst. Henri Poincar\`e \textbf{22}, 229 (1975).
\bibitem[Tsagas (2001)]{Ts1} Tsagas C.G., Phys. Rev. Lett. \textbf{86}, 5421 (2001).
\bibitem[Tsagas (2005)]{Ts2} Tsagas C.G., Class. Quantum Grav. \textbf{22}, 393 (2005).
\bibitem[Tsagas (2006)]{Ts3} Tsagas C.G., Class. Quantum Grav. \textbf{23}, 4323 (2006).
\bibitem[Tsagas, Challinor \& Maartens (2008)]{TCM} Tsagas C.G., Challinor A. and Maartens R., Phys. Rep. \textbf{465}, 61 (2008).
\bibitem[Tsagas \& Kadiltzoglou (2013)]{TK} Tsagas C.G. and Kadiltzoglou M., Phys. Rev. D \textbf{88}, 083501 (2013),
\bibitem[Tsamparlis \& Mason (1983)]{TM} Tsamparlis M. and Mason D.P., J. Math. Phys. \textbf{24}, 157 (1983).
\bibitem[Thorne (1965a)]{Th1} Thorne K.S., Phys. Rev. B \textbf{138}, 251 (1965).
\bibitem[Thorne (1965b)]{Th2} Thorne K.S., Phys. Rev. B \textbf{139}, 244 (1965).
\bibitem[Wald (1984)]{W} Wald R.M., \textit{General Relativity} (University of Chicago Press, Chicago, 1984).
\bibitem[Zafiris (1997)]{Z} Zafiris E., J. Math. Phys. \textbf{38}, 5854 (1997).
\end{thebibliography}
\end{document}